\documentclass[pra, twocolumn, amscd, amsmath, amssymb,letterpaper,showpacs,biblatex,superscriptaddress]{revtex4-1}

\usepackage{graphicx}   % need for figures
\usepackage{epstopdf}   % need for including .eps format figures
\usepackage{verbatim}   % useful for program listings
\usepackage{color}      % use if color is used in text
\usepackage{subfigure}  % use for side-by-side figures
\usepackage{hyperref}   % use for hypertext links, including those to external documents and URLs
\usepackage{bm}         % need for bold math
\usepackage{amsfonts}
\usepackage{eurosym}
\usepackage{amsmath, amsthm, amssymb}
\usepackage{tabularx,colortbl}
\usepackage{dcolumn}    % need for tables
\usepackage{array}
\usepackage{amscd}
\usepackage{dsfont}
\usepackage{bbold}

\newcommand{\ket}[1]{\ensuremath{|{#1}\rangle}}

\newcommand{\red}[1]{\textcolor{black}{#1}}
\newcommand{\up}{\uparrow}
\newcommand{\dn}{\downarrow}

\begin{document}

\title{Characterizing gate operations near the sweet spot of an exchange-only qubit}
\author{Jianjia Fei}
% \email{jfei@wisc.edu}
\affiliation{Department of Physics, University of Wisconsin-Madison, Madison, Wisconsin 53706, USA}
\author{Jo-Tzu Hung}
\affiliation{Department of Physics, University at Buffalo, State University of New York, Buffalo, New York 14260, USA}
\author{Teck Seng Koh}
\affiliation{Department of Physics, University of Wisconsin-Madison, Madison, Wisconsin 53706, USA}
\author{Yun-Pil Shim}
\affiliation{Laboratory for Physical Sciences, College Park, Maryland 20740, USA}
\author{S. N. Coppersmith}
\affiliation{Department of Physics, University of Wisconsin-Madison, Madison, Wisconsin 53706, USA}
\author{Xuedong Hu}
\affiliation{Department of Physics, University at Buffalo, State University of New York, Buffalo, New York 14260, USA}
\author{Mark Friesen}
\affiliation{Department of Physics, University of Wisconsin-Madison, Madison, Wisconsin 53706, USA}
\email{friesen@physics.wisc.edu}
\date{\today} 

\begin{abstract}
Optimal working points or ``sweet spots" have arisen as an important tool for mitigating charge noise in quantum dot logical spin qubits.
\red{The exchange-only qubit provides an ideal system for studying this effect because $Z$ rotations are performed directly at the sweet spot, while $X$ rotations are not.
Here for the first time we quantify the ability of the sweet spot to mitigate charge noise by treating $X$ and $Z$ rotations on an equal footing.
Specifically, we optimize $X$ rotations and determine an upper bound on their fidelity.
We find that sweet spots offer a fidelity improvement factor of at least 20 for typical GaAs devices, and more for Si devices.}
\end{abstract}

\pacs{}
\maketitle

\begin{figure}[t]
\centering
\includegraphics[scale=0.25]{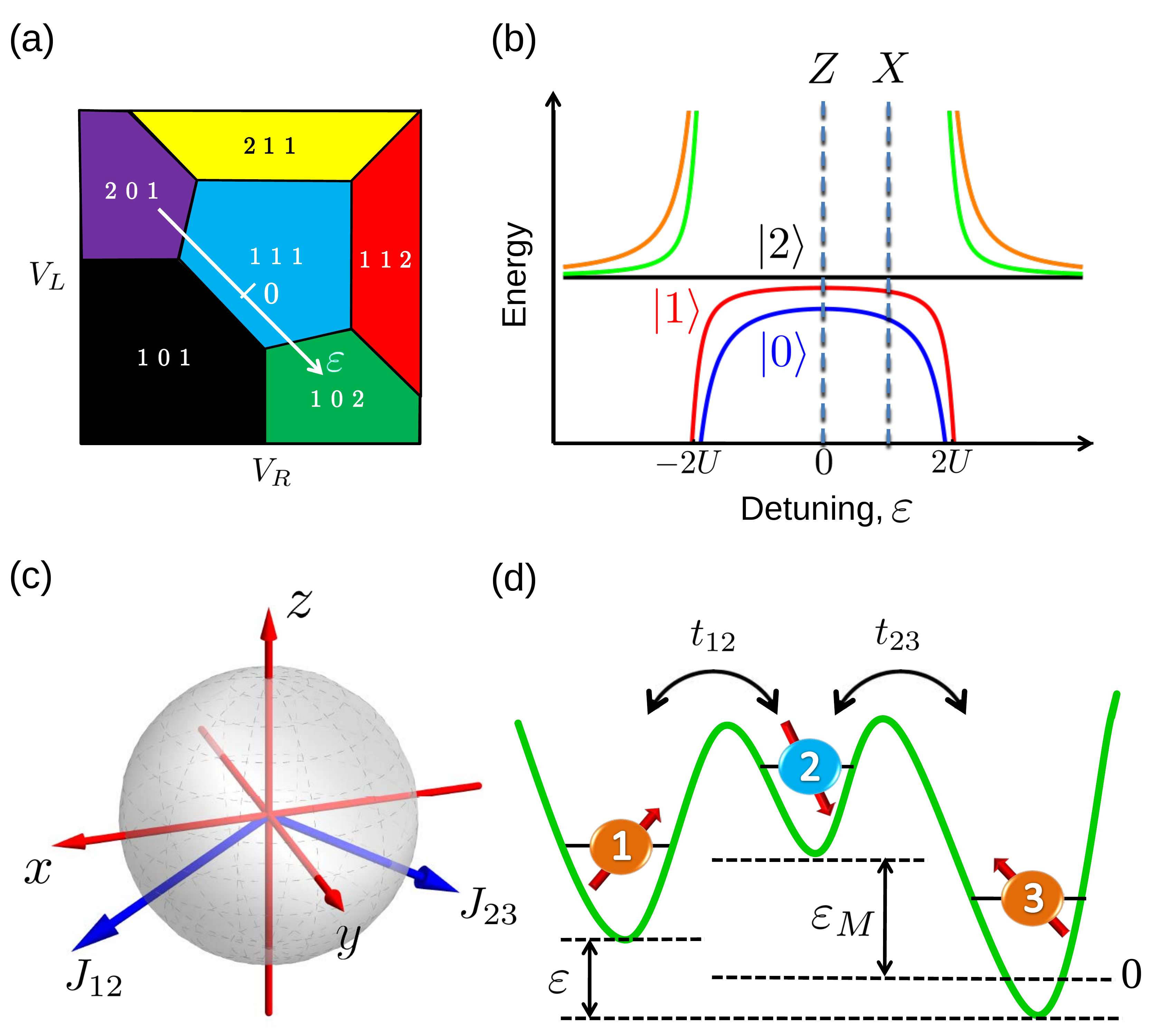}
\caption{(Color online). 
(a) A two-dimensional cut through the charge stability diagram of a triple quantum dot as a function of top-gate voltages, for a fixed value of the detuning parameter $\varepsilon_M$. 
(b) Energy level diagram of the $S_z=1/2$ manifold as a function of the detuning parameter $\varepsilon$. 
In the central region, the low energy states $\ket{0}$-$\ket{2}$ are in the (1,1,1) charge configuration, while the high energy states are doubly occupied.
(c) Bloch sphere representation of the logical qubit, with the rotation axes corresponding to $J_{23}=0$ (left) and $J_{12}=0$ (right). 
(d) Hubbard model of a triple quantum dot containing three electrons.}
\label{fig:SEBH}
\end{figure}

A great challenge in quantum computation is to perform prescribed operations with very small error rates.  Logical qubits are important for achieving this~\cite{Laflamme1996}, since they are fundamental for quantum error correction~\cite{Shor1995}.
Moreover, logical qubits can have symmetries that give rise to so-called sweet spots, at which the effects of noise are suppressed~\cite{Vion2002}.
Several logical spin qubits have been proposed for quantum dot architectures~\cite{Loss1998}.
Here, we consider the exchange-only logical qubit~\cite{DiVincenzo2000}, formed of three electrons in a triple dot~\cite{Gaudreau2006,Laird2010,Gaudreau2011}, as illustrated in Fig.~\ref{fig:SEBH}(d).  
This qubit has the advantage that it has the potential to be very fast, since all operations can be implemented without spatially varying magnetic fields.

The effects of charge noise can never be fully suppressed, even near a sweet spot~\cite{Makhlin2004}.
In this paper, we quantify the effect of sweet spots on gate fidelities by performing theoretical simulations of pulsed gate operations in an exchange-only qubit.
The sweet spot in this device occurs at the symmetry point shown in Fig.~\ref{fig:SEBH}(b), where the detuning parameters $\varepsilon=\varepsilon_M=0$, and the charge-induced fluctuations of the detuning~\cite{Coish2005} are suppressed, to leading order.
(Charge noise in the tunnel coupling~\cite{Hu2006} is not suppressed at this point, but is not thought to be a dominant noise source~\cite{Dial2013}.)
As consistent with recent experiments~\cite{Medford2012}, $Z$-rotations are performed at the sweet spot, while $X$-rotations are obtained by pulsing away from this point.
In principle, the different rotations can be turned on and off independently.
However, in practice it may be necessary to turn on the exchange interactions and magnetic field at all times, to suppress leakage into the non-logical sector of the Hilbert space~\cite{Taylor2013,Hung2014}.

The exchange-only qubit provides an ideal platform for assessing the effect of sweet spots, since all gate operations are generated by the same physical process (the exchange interaction~\cite{DiVincenzo2000,Gimenez2009,Hsieh2012}).
The only difference between $X$ and $Z$-rotations is their proximity to the sweet spot.
The fidelities of these operations can therefore be used to quantify the effectiveness of the sweet spot for mitigating charge noise.
This is in contrast with logical qubits where the different rotation axes correspond to different physical processes (e.g., exchange vs.\ magnetic couplings in singlet-triplet qubits~\cite{Taylor2005,Petta2005}).

In Ref.~\cite{Hung2014}, we provided a detailed account of magnetic noise from the Overhauser fields of nuclear spins on the decoherence of an exchange-only qubit.
Here, we simulate realistic gate operations including quasistatic random Overhauser fields~\cite{Mehl2013} and charge noise~\cite{Taylor2013}.
In certain regimes we find that the main limit on the gate fidelities arises from the Overhauser fields, as consistent with experimental observations~\cite{Medford2012}.
However, when the gates are properly optimized, we predict that charge noise should determine the upper bound on gate fidelities.
After optimization, we find that gate fidelities at the sweet spot are typically 20 times better than away from the sweet spot.

\emph{Theoretical Model.}---
We model the coherent evolution of the exchange-only qubit using a 3-electron, 3-site Hubbard model with the Hamiltonian~\cite{SI}
\begin{align}
\label{eq:Hamiltonian}
H=&\sum_{\langle i,j\rangle\sigma} t_{ij}(c_{i\sigma}^\dagger c_{j\sigma}+c_{j\sigma}^\dagger c_{i\sigma})+U\sum_{j}n_{j\uparrow}n_{j\downarrow}\notag\\
  &+\sum_{j}\varepsilon_j(n_{j\uparrow}+n_{j\downarrow})-g\mu_B B\sum_j (n_{j\uparrow}-n_{j\downarrow})\notag\\
  &+g\mu_B\sum_j \Delta B_j(n_{j\uparrow}-n_{j\downarrow}),
\end{align}
where the labels $\{ i,j\} =1,2,3$ correspond to dot locations, $\uparrow,\downarrow$, and $\sigma$ refer to individual spin $s_z$ eigenstates, $c_{j\sigma}^\dagger$ and $c_{j\sigma}$ are electron creation and annihilation operators, and $n_{j\sigma}$ is the electron number operator.
The first term in Eq.~(\ref{eq:Hamiltonian}) describes the tunneling, with tunnel couplings $t_{ij}$.
We assume a symmetric, linear triple dot geometry, as shown in Fig.~\ref{fig:SEBH}(d), with $t_{12}=t_{23}\equiv t$ and $t_{13}=0$.
The second term describes the onsite Coulomb repulsion, with energies $U$ that are the same at every site.
\red{
(Off-site Coulomb interactions are discussed in~\cite{SI}.)}
The third term describes the local electrostatic potentials $\varepsilon_j$.
The fourth term describes the Zeeman energy due to a uniform external magnetic field ${\bf B}=B\hat{\bf z}$, with the Land\'{e} $g$-factor and Bohr magneton $\mu_B$.
The fifth term describes the local variations of the Zeeman energy due to Overhauser field fluctuations $\Delta {\bf B}_j$.
Here, we take $\Delta {\bf B}_j\| {\bf B}$ because the lateral components of $\Delta {\bf B}_j$ generate couplings between $S_z$ manifolds that are highly suppressed in the regime of large Zeeman splittings, which we consider below.
We also ignore Coulomb interactions between electrons in different dots.
The detuning $\varepsilon=\varepsilon_1-\varepsilon_3$ is defined in analogy with experiments~\cite{Laird2010, Medford2012}, and corresponds to the energy difference between the (2,0,1) and (1,0,2) charge configurations. 
For a triple dot, there is also a second, independent detuning parameter~\cite{Hickman2013}, which we define here as $\varepsilon_M=\varepsilon_2-(\varepsilon_1+\varepsilon_3)/2$.
In experimental systems, the detunings $\varepsilon$ and $\varepsilon_M$ are controlled by voltages, including $V_L$ and $V_R$, which are applied to the top-gates.
A typical charge stability diagram is shown in Fig.~\ref{fig:SEBH}(a) for a fixed value of $\varepsilon_M$.

The Hilbert space associated with Eq.~(\ref{eq:Hamiltonian}) is large.
For GaAs-based devices, most leakage channels can be suppressed by enforcing sizeable energy splittings~\cite{Hung2014}.
As consistent with recent experiments~\cite{Medford2012}, we therefore consider the energy hierarchy 
$g\mu_BB\gg J\gg g\mu_B\Delta B>0$, where $J$ is the exchange interaction generated by the tunnel couplings.
(In \cite{SI}, we briefly consider $^{28}$Si-based devices, which do not require such an energy heirarchy, due to the absence of nuclear spins.)
Since $g \mu_B B$ is large, the energy spectrum splits into manifolds of constant total spin $S_z$.  
Our simulations focus on the seven states in the $S_z=1/2$ manifold, where the two qubit states are defined in the decoherence free subspace with $S=S_z$~\cite{DiVincenzo2000,Medford2012}.  
For a basis set, we consider the seven eigenstates of Eq.~(\ref{eq:Hamiltonian}) when $\varepsilon=\varepsilon_M=\Delta B_j=0$, consisting of three singly-occupied $(1,1,1)$ states,
$\vert0\rangle = \sqrt{1/3}\vert T_0\rangle_{13}\ket{\uparrow}_2-\sqrt{2/3}\vert T_+\rangle_{13}\ket{\downarrow}_2$,
$\vert1\rangle = \vert S\rangle_{13}\ket{\uparrow}_2$, and
$\vert 2\rangle = \sqrt{2/3}\vert T_0\rangle_{13}\ket{\uparrow}_2+\sqrt{1/3}\vert T_+\rangle_{13}\ket{\downarrow}_2$,
and four doubly-occupied states,
$\vert 3\rangle=\vert S\rangle_1\vert\cdot\rangle_2\ket{\uparrow}_3$,
$\vert 4\rangle=\vert\cdot\rangle_1\vert S\rangle_2\ket{\uparrow}_3$,
$\vert 5\rangle=\ket{\uparrow}_1\vert S\rangle_2\vert\cdot\rangle_3$,
and $\vert 6\rangle=\ket{\uparrow}_1\vert\cdot\rangle_2\vert S\rangle_3$.
Here, the subscript denotes the dot index, $\vert S\rangle = \frac{1}{\sqrt{2}}(\ket{\uparrow\downarrow} - \ket{\downarrow\uparrow})$, $\vert T_0\rangle = \frac{1}{\sqrt{2}}(\ket{\uparrow\downarrow} + \ket{\downarrow\uparrow})$, and $\vert T_+\rangle = \ket{\uparrow\uparrow}$ are the singlet and triplet states of two spins, and $\vert\cdot\rangle$ represents a dot with no electrons.
$\vert0\rangle$ and $\vert1\rangle$ are the logical qubit states, $|2\rangle$ is the main leakage state, and the doubly occupied states mediate the exchange interaction.

The gate simulations described below include the full set of seven basis states, in order to address questions of leakage and decoherence.
However it is instructive to consider the effective Hamiltonian in the $\{\ket{0},\ket{1}\}$ logical subspace~\cite{Taylor2013,SI},
\begin{equation}
H = \frac{\sqrt{3}}{4}(J_{12}-J_{23})\sigma_x-\frac{1}{4}(J_{12}+J_{23})\sigma_z ,
\label{eq:H2}
\end{equation}
where $J_{12}$ and $J_{23}$ are exchange interactions.
The latter may be tuned independently as a function of the control parameters $\varepsilon$ and $\varepsilon_M$, yielding a continuous set of rotations in the $x$-$z$ plane of the Bloch sphere.
For example, we could independently set $J_{12}$ or $J_{23}$ to zero, yielding the pair of rotation axes shown in Fig.~\ref{fig:SEBH}(c).
From Eq.~(\ref{eq:H2}), we see that $Z$-rotations are obtained when $J_{12}=J_{23}$.
In \cite{SI}, we show that this requirement is met when either $\varepsilon=0$ or $\varepsilon_M=0$.
We also show that the special combination $\varepsilon=\varepsilon_M=0$ corresponds to a detuning sweet spot, because 
$\partial E_{01}/\partial \varepsilon =\partial E_{01}/\partial \varepsilon_M=0$, where $E_{01}$ is the energy splitting between the qubit states.
Since always-on exchange interactions are needed to prevent leakage, and since $J_{12},J_{13}>0$, Eq.~(\ref{eq:H2}) suggests that we cannot achieve pure $X$-rotations.
We overcome this problem by implementing a three-step pulse sequence~\cite{Hanson2007PRL}.
This procedure requires moving away from the sweet spot, with consequences for the decoherence and gate fidelity.
Finally, we note that a complete set of single-qubit operations must include initialization and readout. 
The latter are accomplished in experiments by adiabatically tuning the device to the $(2,0,1)$ or $(1,0,2)$ charge configurations in the far-detuned regime of Fig.~\ref{fig:SEBH}(b)~\cite{Medford2012}.
In our simulations, we do not investigate readout and initialization; we consider only the unitary gate operations.
Moreover, we assume instantaneous (diabatic) pulses and do not investigate pulse imperfections.
We consider only the errors caused by charge and nuclear noise sources, and by leakage outside the logical qubit Hilbert space.

\emph{Gate Simulations.}---
We simulate the dynamics of the logical qubit gate operations by solving the master equation
\begin{equation}
\label{eq:master}
\frac{d\rho(t)}{dt} = -\frac{i}{\hbar}[H,\rho(t)]-D[\rho(t)]
\end{equation}
for the $7\times 7$ density matrix, $\rho$.
The first term on the right-hand side of Eq.~(\ref{eq:master}) describes the unitary evolution, while the second term describes the decoherence.

We consider dephasing from charge noise and random Overhauser fields.
The nuclear fluctuations occur at frequencies much lower than the relevant electronic time scales~\cite{Taylor2007};
we take them to be quasistatic with a Gaussian distribution of width $\sigma_B=4$~mT, as appropriate for GaAs~\cite{Assali2011}.
\red{
We model the charge noise as either much faster than the qubit gate frequency, with a Markovian dephasing rate of $\Gamma \sim 1$~GHz~\cite{Hayashi2003}, or much slower than the qubit frequency, with a Gaussian distribution of width $\sigma_\varepsilon=5$~$\mu$eV~\cite{Petersson2010,Shi2012}.
Both noise models have been invoked previously to desribe charge noise in similar scenarios~\cite{Taylor2007,Barrett2002}.
We do not specifically treat noise at the qubit rotation frequency, which, in contrast to resonantly driven systems~\cite{Jing2014}, does not play a special role in determining the fidelity of dc pulsed gates.
In our Markovian model, all high frequency noise is treated as uncorrelated.}

We consider two types of fast charge noise.
The virtually occupied states $\ket{3}$-$\ket{7}$ mediate exchange interactions, but they also contribute to double occupation dephasing errors of the form~\cite{Barrett2002} $D_U=\sum_i \frac{\Gamma}{2}[n_{i\uparrow}+n_{i\downarrow},[n_{i\uparrow}+n_{i\downarrow},\rho]]$. 
We also consider direct dephasing $D_\varepsilon$ of the singly occupied states $\ket{0}$-$\ket{2}$, with rates that depend on the derivative of the energy splitting $E_{ij}$ between eigenstates $\ket{i}$ and $\ket{j}$ with respect to the detuning~\cite{Taylor2007}.
We assume that contributions from the individual detuning parameters contribute in quadrature, with the dephasing rates $\gamma_{ij}=\Gamma [(\partial E_{ij}/\partial \varepsilon)^2+2(\partial E_{ij}/\partial \varepsilon_M)^2]^{1/2}$.
Here, the factor of 2 reflects the relative magnitudes of the $\varepsilon$ and $\varepsilon_M$ terms in the effective $2\times 2$ Hamiltonian for the logical qubit states~\cite{SI}.
The resulting dephasing matrix is given by
\begin{align}
& D[\rho(t)] = D_U + D_\varepsilon = \\
& \hspace{-0.05in} \begin{pmatrix}
0 & \gamma_{01}\rho_{01} & \gamma_{02}\rho_{02} & \Gamma\rho_{03} & \Gamma\rho_{04} &\Gamma\rho_{05} & \Gamma\rho_{06} \\
\gamma_{01}\rho_{01}^* & 0 & \gamma_{12}\rho_{12} & \Gamma\rho_{13} & \Gamma\rho_{14} & \Gamma\rho_{15} & \Gamma\rho_{16}\\
\gamma_{02}\rho_{02}^* & \gamma_{12}\rho_{12}^* & 0 & \Gamma\rho_{23} & \Gamma\rho_{24} & \Gamma\rho_{25} & \Gamma\rho_{26}\\
\Gamma\rho_{03}^* & \Gamma\rho_{13}^* & \Gamma\rho_{23}^* & 0 & 4\Gamma\rho_{34} & 3\Gamma\rho_{35} & \Gamma\rho_{36}\\
\Gamma\rho_{04}^* & \Gamma\rho_{14}^* & \Gamma\rho_{24}^* & 4\Gamma\rho_{34}^* & 0 & \Gamma\rho_{45} & 3\Gamma\rho_{46}\\
\Gamma\rho_{05}^* & \Gamma\rho_{15}^* & \Gamma\rho_{25}^* & 3\Gamma\rho_{35}^* & \Gamma\rho_{45}^* & 0 & 4\Gamma\rho_{56}\\
\Gamma\rho_{06}^* & \Gamma\rho_{16}^* & \Gamma\rho_{26}^* & \Gamma\rho_{36}^* & 3\Gamma\rho_{46}^* & 4\Gamma\rho_{56}^* & 0 
\end{pmatrix} . \nonumber
\end{align}

We treat the slow fluctuations of the detuning and Overhauser fields by numerically solving the 49 coupled real differential equations in Eq.~(\ref{eq:master}) for a fixed noise realization~\cite{SI}.
We then repeat the calculations for 625 realizations of Overhauser field fluctuations and 961 realizations of detuning fluctuations, and perform the appropriate Gaussian averages.  
The simulations are performed on the Open Science Grid at the University of Wisconsin-Madison~\cite{OSG}.
The results reported here represent $>23$ compute years.

\emph{Gate Optimization.}---
We begin by considering $Z(\pi)$ rotations of the logical qubit.
As described above, these operations are performed at the sweet spot $\varepsilon=\varepsilon_M=0$.
Fluctuations of the detuning and the Overhauser fields give rise to errors within the qubit subspace as well as leakage.
We monitor these effects by performing quantum process tomography (QPT)~\cite{SI}, beginning the simulations in four different initial states, and comparing the final results to the ideal final states for a fixed value of the tunnel coupling $t$.
In this procedure, the evolution period $\tau$ is treated as a variable.
The optimal value of $\tau$ is chosen by maximizing the fidelity $F$ obtained from QPT, with results shown in Fig.~2(a).
For small $t$, the rotations are slow, and the fidelity is strongly suppressed by the quasistatic random Overhauser fields.
For large $t$, the rotations are fast, and the fidelity is determined by a combination of charge noise and leakage.
Since the leakage process is coherent, the projection of the full density matrix onto the logical qubit subspace undergoes oscillations, as seen in the lower inset of Fig.~2(a).
These oscillations are severe for large tunnel couplings, causing a deterioration of the fidelity as seen in the main figure.

We also investigate $X(\pi)$ rotations of the logical qubit.
As noted above, it is not possible to perform a direct rotation around $\hat{\bf x}$; accurate rotations require multi-pulse gate sequences.
Here, we consider a three-step procedure~\cite{Hanson2007PRL} that can be visualized as shown in the lower inset of Fig.~2(c).
The sequence consists of (i) a $\pi$-rotation around the $-(\hat{\bf x}+\hat{\bf z})/\sqrt{2}$ axis on the Bloch sphere, 
(ii) a $Z(\pi)$ rotation, 
and (iii) a final $\pi$-rotation around the $-(\hat{\bf x}+\hat{\bf z})/\sqrt{2}$ axis.
For steps (i) and (iii), the values of $\varepsilon$ and $\varepsilon_M$ that determine the axis tilt are not known \emph{a priori}; we find them by performing fidelity simulations for the desired gate operations in the absence of detuning and nuclear fluctuations.
The results are shown in Fig.~2(b) for a fixed value of $t$, \red{with a line shape that is analytically determined in~\cite{SI}.
The optimal fidelities along this line are obtained via simulations, as indicated by the red star.}
This calibration procedure is then repeated for other values of $t$.
The three-step protocol is then optimized, step by step, by performing simulations to determine the evolution period $\tau$ that maximizes the fidelity of each step.
The final fidelities of the three-step $X(\pi)$ protocol are shown in Fig.~2(c).
We observe results similar to those in Fig.~2(a).
However, the effects of leakage and charge noise are more severe because steps~(i) and (iii) are not performed at sweet spots.
The suppression of the fidelity due to leakage is most obvious at large $t$.
The lower inset shows a typical evolution projected onto the logical qubit Bloch sphere.

\begin{figure*}[th]
\centering
\includegraphics[width=6in]{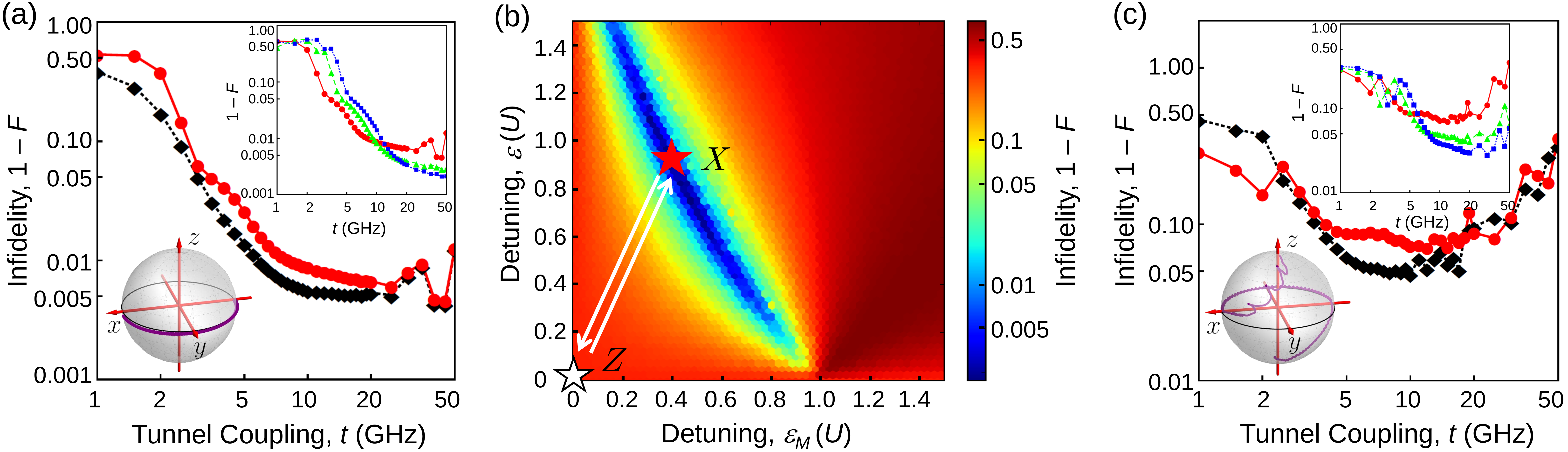}
\caption{(Color online). 
Optimized rotation fidelities, $F$, obtained from simulations of the exchange-only qubit, including random nuclear fields and detuning fluctuations typical of GaAs quantum dots.
(a) $Z(\pi)$ rotations.
Circles include fast and slow detuning fluctuations, while the longitudinal magnetic field gradients are held fixed at $\Delta B_2^z-\Delta B_1^z=\Delta B_3^z-\Delta B_2^z =3$~mT.
Diamonds include quasistatic fluctuations of the Overhauser fields and fast detuning noise, but no slow detuning fluctuations.
Both solutions assume an onsite Coulomb repulsion of $U=1$~meV.  
For small $t$, the fidelity is mainly limited by nuclear noise and leakage into state $\ket{2}$, while for large $t$, the fidelity plateau is mainly limited by charge noise.
For very large $t$, leakage into the excited charge states causes fidelity oscillations that are nearly independent of nuclear noise~\cite{SI}.
The lower inset shows the evolution of the density matrix projected onto the Bloch sphere of the logical qubit for the tunnel coupling $t=10$~GHz; the small, rapid oscillations are caused by leakage.
The upper inset shows results of averaging over detuning noise for $U=1$~meV (circles, as in the main figure), $U=2$~meV (triangles), and $U=3$~meV (squares), with larger $U$ yielding higher maximum fidelities. 
(b) Fidelity of $\pi$-rotations around the axis $-(\hat{\bf x}+\hat{\bf z})/\sqrt{2}$, in the absence of noise, corresponding to step~(i) of a three-step $X(\pi)$ rotation protocol \cite{Hanson2007PRL}, for $t=5$~GHz.
The red star indicates the optimal values of $\varepsilon$ and $\varepsilon_M$.
(c) Final fidelity of $X(\pi)$ rotations, via the three-step protocol, where step~(i) occurs at the red star in (b), step~(ii) occurs at the white star ($\varepsilon=\varepsilon_M=0$), and step~(iii) occurs at the red star.
The circles and diamonds have the same meaning as in (a).
Here, the fidelity-limiting mechanisms are similar to (a), with a much stronger suppression of the fidelity at large $t$, due to leakage and charge noise.
The insets are also defined as in (a).
Note the large leakage oscillations during steps (i) and (iii) of the protocol.}
\label{fig:ZX_InFid_depsilon}
\end{figure*}

\emph{Results and Discussion.}---
The $X$ and $Z$-rotation protocols used in Fig.~2 are different.
However, by comparing fidelities obtained using QPT, we can compare the final results effectively.
We observe that maximal fidelities (or minimal infidelities, $1-F$) occur over a range of moderate to large tunnel couplings, $t\simeq 5$-20~GHz, that depends on the Hubbard repulsion parameter $U$.
Our results also depend on the local field gradients $\Delta B_j$, which determine the leakage rate.
The values of $\Delta B_j$ considered here are typical for GaAs triple dots.
The optimal fidelities in Fig.~2 occur on a plateau, whose value is largely determined by the detuning noise.  
This is not the same conclusion reached in \cite{Medford2012}, where fidelity limits were attributed to nuclear noise.
We speculate that those experiments were performed at lower $t$, below the plateau, where nuclear noise predominates.
We emphasize that larger $t$ should be used to achieve maximal fidelities.

Our most important results are obtained by comparing the maximal fidelities of $X(\pi)$ and $Z(\pi)$ rotations.
We find that $X$-rotations have maximum fidelities $\sim$20 times worse than $Z$-rotations, which can be directly attributed to the fact that $X$-rotations occur away from the sweet spot, while $Z$-rotations occur at the sweet spot.
The degradation of $X$-rotations is most noticeable for large $t$, where the fidelity is dominated by charge noise.
Measurements of the quantum dot hybrid qubit show a similar degradation of coherence away from a sweet spot~\cite{Shi2014,Kim2014}.
This suggests that AC gating techniques could yield better fidelities than the DC pulsing techniques studied here, because the detuning is always centered at the sweet spot~\cite{Taylor2013}.
Indeed, recent experiments on the exchange-only qubit have employed such a strategy~\cite{Medford2013}.
On the other hand, AC methods tend to produce somewhat slower gates, for which nuclear noise could be a problem.

Finally, we note that our analysis has focused on GaAs quantum dot devices, where nuclear noise is known to be important. 
For Si-based devices, especially isotopically purified $^{28}$Si, the nuclear noise can be very small. 
As a result, Si devices should yield better fidelities for DC pulsed gates, especially in the low-$t$ regime.
(See \cite{SI}.) 
At higher $t$, where nuclear noise is not predominant, Si and GaAs exchange-only qubits should have similar fidelities.

We thank Z. Shi, X. Wu, K. Rudinger, J. Gamble, and C. Wong for helpful discussions. 
We also thank the HEP, Condor, and CHTC groups at UW-Madison for computational support. 
This work was supported in part by ARO (W911NF0910393), NSF (PHY-1104660 and PHY-1104672), UW-Madison (150 486700 4), and by the United States Department of Defense. 
The views and conclusions contained in this paper are those of the authors and should not be interpreted as representing the official policies, either expressed or implied, of the US Government.

\renewcommand{\theequation}{\Alph{section}\arabic{equation}}
\renewcommand{\thefigure}{S\arabic{figure}}

\begin{appendix}

\setcounter{section}{0}
\setcounter{equation}{0}
\setcounter{figure}{0}

\begin{center}
% \textbf{Supplementary Materials}
\textbf{Appendix}
\end{center}
In these Appendices,
% Supplementary Materials, 
we provide details about the calculations and simulations discussed in the main text.
Section~A describes the Hamiltonian for the $S_z=1/2$ spin manifold.
Section~B provides analytical estimates for the exchange interactions in certain operating regimes of interest.
Section~C provides details of the quantum process tomography methods.
Section~D describes our statistical averaging procedure for treating quasistatic charge and nuclear noise.
Section~E describes some additional results for simulations with averages over Overhauser field gradients. 
Section~F describes results with no Overhauser field gradients, consistent with pure, isotopically purified $^{28}$Si.

\begin{figure*}[t]
\centering
\includegraphics[width=4.5in]{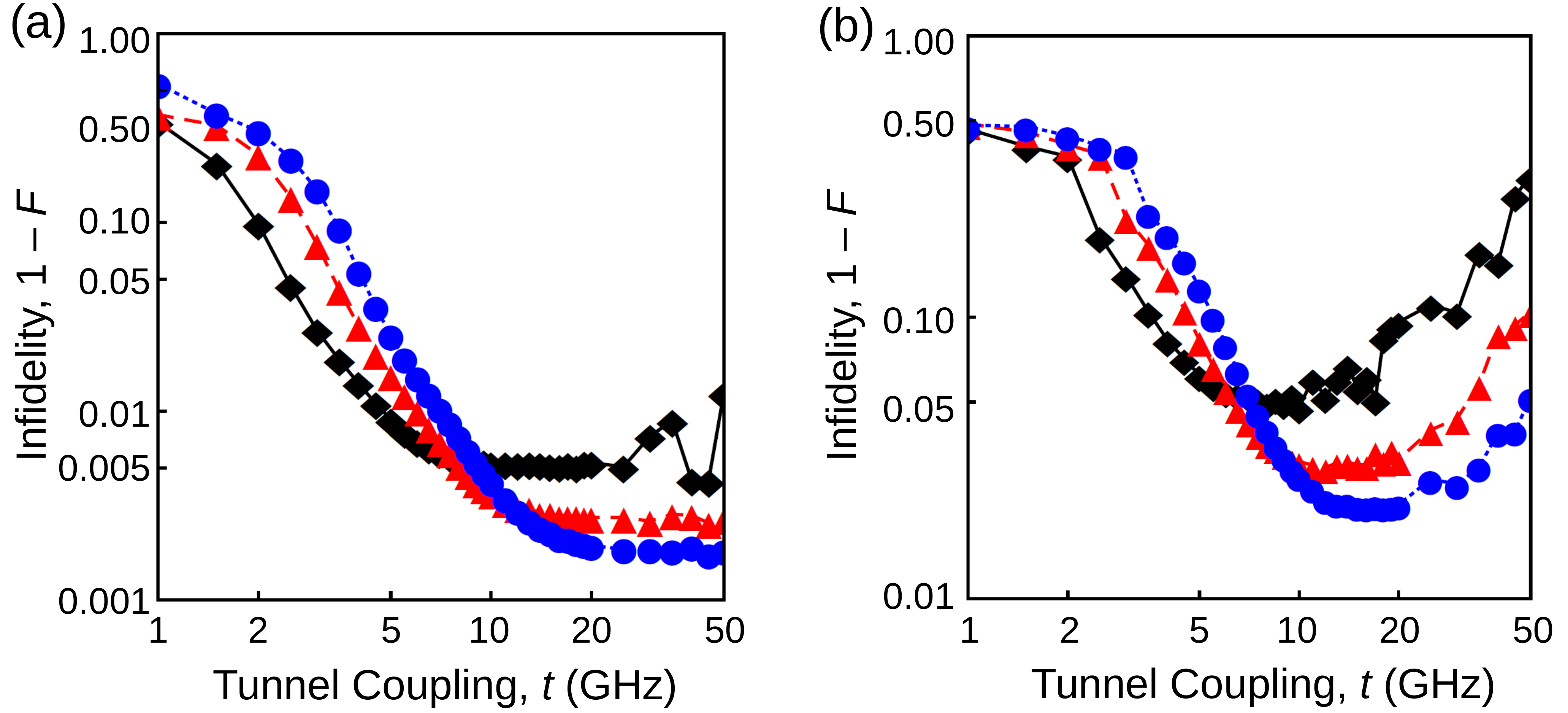}
\caption{
Optimized gate fidelities, $F$, obtained from simulations of the exchange-only qubit.
(a) Infidelity, $1-F$, of a $Z(\pi)$ rotation, including quasistatic nuclear noise for three different values of the intradot Coulomb repulsion: $U=1$~meV (black diamonds, as in the main panel of Fig.~2(a) in the main text), $U=2$~meV (triangles), and $U=3$~meV (circles), with larger $U$ values yielding higher maximum fidelities. 
(b) Fidelity of a three-step $X(\pi)$ rotation, as described in the main text, including quasistatic nuclear noise for three different values of the Coulomb repulsion: $U=1$~meV (black diamonds, as in the main panel of Fig.~2(c) in the main text), $U=2$~meV (triangles), and $U=3$~meV (circles). 
All fidelity averages are obtained assuming a Gaussian distribution of Overhauser field differences with standard deviation $\sigma_B=4$~mT. Quasistatic noise in the detuning parameters are not included in this simulation.} 
\label{fig:ZX_InFid_deltaB}
\end{figure*}

\section{Calculation Details}
In this section, we describe our Hubbard model Hamiltonian.
We evaluate each individual term of Eq.~(1) in the main text using the 7D basis set defined by
\begin{gather}
\ket{0}=\frac{1}{\sqrt{6}}\left(\ket{\up\up\dn}+\ket{\dn\up\up}\right)-\sqrt{\frac{2}{3}}\ket{\up\dn\up} , \\
\ket{1}=\frac{1}{\sqrt{2}}\left(\ket{\up\up\dn}-\ket{\dn\up\up}\right) , \\
\ket{2}=\frac{1}{\sqrt{3}}\left(\ket{\up\up\dn}+\ket{\dn\up\up}\right)+\sqrt{\frac{1}{3}}\ket{\up\dn\up} , \\
\ket{3}=\frac{1}{\sqrt{2}}\left(\ket{\up\dn}_1-\ket{\dn\up}_1\right)\ket{\cdot}_2\ket{\up}_3 , \\
\ket{4}=\frac{1}{\sqrt{2}}\ket{\cdot}_1\left(\ket{\up\dn}_2-\ket{\dn\up}_2\right)\ket{\up}_3 ,  \\
\ket{5}=\frac{1}{\sqrt{2}}\ket{\up}_1\left(\ket{\up\dn}_2-\ket{\dn\up}_2\right)\ket{\cdot}_3 , \\
\ket{6}=\frac{1}{\sqrt{2}}\ket{\up}_1\ket{\cdot}_2\left(\ket{\up\dn}_3-\ket{\dn\up}_3\right) ,
\end{gather}
where the notation $\ket{\up\dn}_j$ (or $\ket{\dn\up}_j$) indicates that both electrons are in the same dot, labelled $j=1,2,3$, and $\ket{\cdot}_j$ indicates an empty dot.
The creation-annihilation operator combinations, 
$c^\dagger_{i\sigma}c_{j\sigma}$, are then readily evaluated, as are the particle number operators $n_{i\sigma}=c^\dagger_{i\sigma}c_{i\sigma}$, for dots $i,j$, and spins $\sigma = \uparrow, \downarrow$.

We then obtain the following expressions for the individual terms in the Hubbard Hamiltonian, Eq.~(1) in the main text.
The tunnel coupling term is given by 

\begin{equation}
H_t = \begin{pmatrix}
0  & 0  & 0  & \sqrt{\frac{3}{2}}t   & -\sqrt{\frac{3}{2}}t  & -\sqrt{\frac{3}{2}}t   & \sqrt{\frac{3}{2}}t \\
0  & 0  & 0  & -\frac{1}{\sqrt{2}}t  & \frac{1}{\sqrt{2}}t   & -\frac{1}{\sqrt{2}}t   & \frac{1}{\sqrt{2}}t \\
0                          & 0                           & 0  & 0  & 0  & 0  & 0\\
\sqrt{\frac{3}{2}}t   & -\frac{1}{\sqrt{2}}t   & 0  & 0  & 0  & 0  & 0\\
-\sqrt{\frac{3}{2}}t  & \frac{1}{\sqrt{2}}t   & 0  & 0  & 0  & 0  & 0\\
-\sqrt{\frac{3}{2}}t  & -\frac{1}{\sqrt{2}}t   & 0  & 0  & 0  & 0  & 0\\
\sqrt{\frac{3}{2}}t   & \frac{1}{\sqrt{2}}t    & 0  & 0  & 0  & 0  & 0
\end{pmatrix}.
\vspace{1in}
\end{equation}

\begin{figure*}[t]
\centering
\includegraphics[width=4.5in]{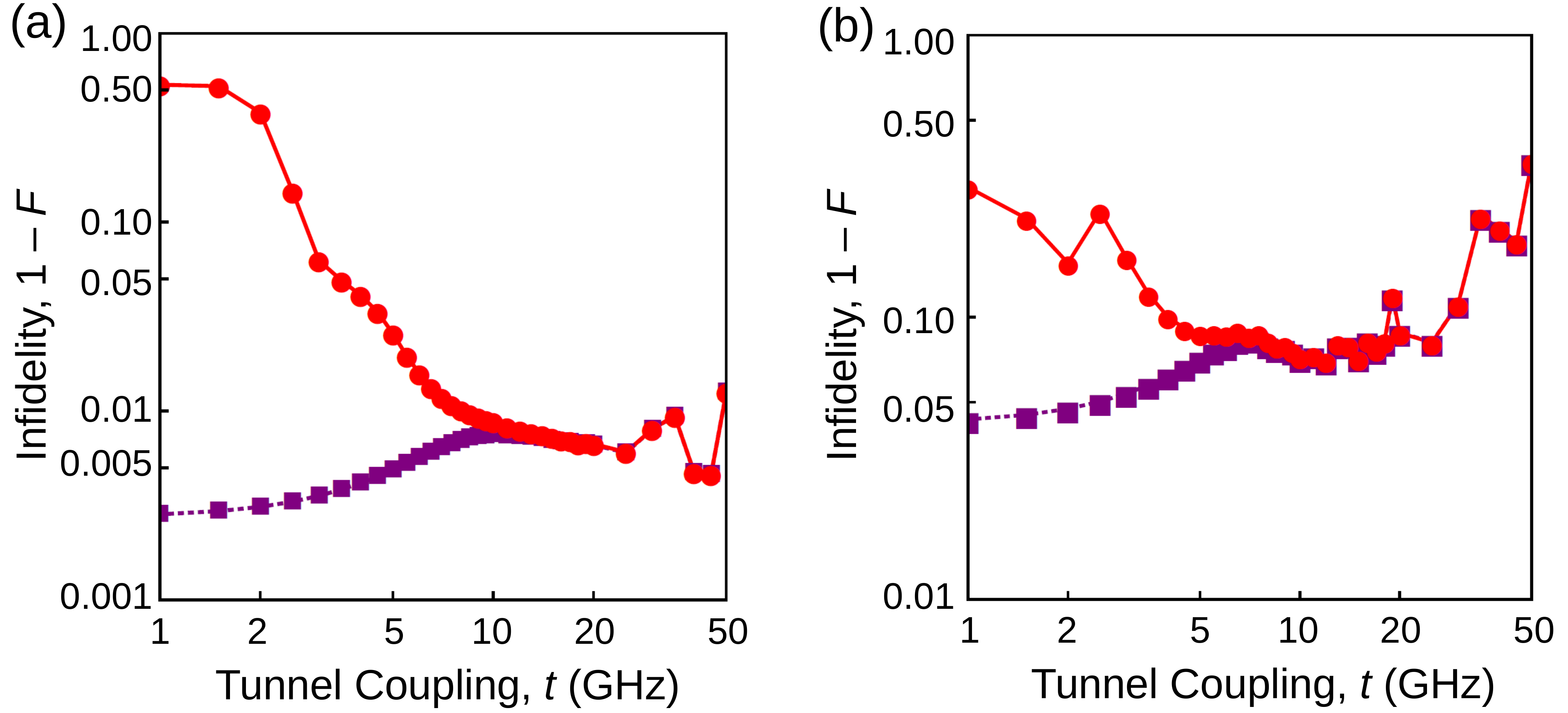}
\caption{ 
Comparison of gate infidelities, $1-F$, for perfect, isotopically purified $^{28}$Si (purple squares) and GaAs (red circles) in the presence of detuning fluctuations.  
The GaAs results are identical to the red circles in Fig.~2 of the main text.
For $^{28}$Si, we assume no Overhauser fields in the dots, while for GaAs, we assume the fixed values $\Delta B_2^z-\Delta B_1^z=\Delta B_3^z-\Delta B_2^z =3$~mT, as for the red circles in Fig.~2.
In both cases, we assume an intradot Coulomb repulsion of $U = 1$~meV. 
(a) $Z(\pi)$-rotations. 
(b) Three-step $X(\pi)$ rotations, as described in the main text.}
\label{fig:ZX_InFid_GaAs_Si}
\end{figure*}

The onsite Coulomb repulsion term describes the double-occupation energy cost for a single dot.
It is given by
\begin{equation}
\vspace{-.8in}
H_U = \begin{pmatrix}
0  & 0  & 0  & 0  & 0  & 0  & 0\\
0  & 0  & 0  & 0  & 0  & 0  & 0\\
0  & 0  & 0  & 0  & 0  & 0  & 0\\
0  & 0  & 0  & U  & 0  & 0  & 0\\
0  & 0  & 0  & 0  & U  & 0  & 0\\
0  & 0  & 0  & 0  & 0  & U  & 0\\
0  & 0  & 0  & 0  & 0  & 0  & U
\end{pmatrix}. \label{eq:HU}
\vspace{.8in}
\end{equation}
\begin{widetext}
The detuning energies are given by 
\begin{equation}
H_\varepsilon = 
\begin{pmatrix}
\varepsilon_M  & 0  & 0  & 0  & 0  & 0  & 0\\
0  & \varepsilon_M  & 0  & 0  & 0  & 0  & 0\\
0  & 0  & \varepsilon_M  & 0  & 0  & 0  & 0\\
0  & 0  & 0  & \frac{\varepsilon}{2}  & 0   & 0  & 0\\
0  & 0  & 0  & 0  & 2\varepsilon_M-\frac{\varepsilon}{2}   & 0  & 0\\
0  & 0  & 0  & 0  & 0  & 2\varepsilon_M+\frac{\varepsilon}{2}   & 0\\
0  & 0  & 0  & 0  & 0  & 0  & -\frac{\varepsilon}{2}
\end{pmatrix}.
\end{equation}
Since the basis states all belong to the same $S_z = 1/2$ spin manifold, they all have the same Zeeman energy, which we ignore here.
The local Overhauser field energies due to nuclear fluctuations are given by
\begin{equation}
H_{\Delta B} = g\mu_B
\begin{pmatrix}
\frac{2}{3}(\Delta B_l - \Delta B_r)  & \frac{1}{\sqrt{3}}(\Delta B_l + \Delta B_r)  & -\frac{\sqrt{2}}{3}(\Delta B_l - \Delta B_r)  & 0   & 0  & 0   & 0\\
\frac{1}{\sqrt{3}}(\Delta B_l + \Delta B_r)  & 0  & \sqrt{\frac{2}{3}}(\Delta B_l + \Delta B_r)  & 0  & 0  & 0   & 0\\
-\frac{\sqrt{2}}{3}(\Delta B_l - \Delta B_r)  & \sqrt{\frac{2}{3}}(\Delta B_l + \Delta B_r) & \frac{1}{3}(\Delta B_l - \Delta B_r)  & 0   & 0  & 0   & 0\\
0  & 0  & 0  & -\Delta B_r  & 0   & 0   & 0\\
0  & 0  & 0  & 0   & -\Delta B_r  & 0   & 0\\
0  & 0  & 0  & 0   & 0    & \Delta B_l  & 0\\
0  & 0  & 0  & 0   & 0    & 0  & \Delta B_l
\end{pmatrix},
\end{equation}
\end{widetext}
where we define $\Delta B_l = B_1^z - B_2^z$ and $\Delta B_r = B_2^z - B_3^z$ to be the differences in local magnetic fields in the $\hat{\bf z}$ direction.
As explained in the main text, we only consider longitudinal ($\hat{\bf z}$) components of the Overhauser fields, defined by ${\bf B}=B\hat{\bf z}$, as consistent with \cite{Taylor2007}.

Equation (3) of the main text then represents a set of 49 real coupled differential equations.
We solve these equations numerically and check that the trace-preserving condition is satisfied for the final density matrix, when the calculation is complete.

\red{
To complete this section, we note that the single-parameter model of Coulomb interactions in Eq.~(\ref{eq:HU}) was chosen for simplicity.
In a more elaborate model, we could expand this to include the following terms
\begin{equation}
U_0\sum_{j=1,2,3}n_{j\uparrow}n_{j_\downarrow}+U_1\sum_{j=1,2}n_jn_{j+1}+U_2\,n_1n_3 ,
\end{equation}
corresponding to double-occupations, nearest-neighbor couplings, and next-nearest-neighbor couplings.
Such models have previously been explored by some authors~\cite{Taylor2013,Gaudreau2006,Korkusinski07}, while other authors~\cite{DiVincenzo2000} consider a basis set of singly-occupied states, where all the Coulomb terms are incorporated into the exchange interaction parameters of an effective Hamiltonian.}

\red{
We can estimate the relative magnitudes of the $U_1$ and $U_2$ terms, which are not included in Eq.~(\ref{eq:HU}), by assuming the following probability density for the lateral distribution of the electronic wavefunctions:
\begin{equation}
|\psi(x,y)|^2 = \frac{1}{2\pi R^2} e^{-(x^2+y^2)/2R^2} .
\end{equation}
To leading order in the ratio $R/L$, where $R$ represents the lateral size of a dot, and $L$ is the interdot separation, we obtain the following relations between the Coulomb interactions:
\begin{gather}
U_0=\sqrt{\pi}\frac{L}{R}U_1 , \\
U_2=\frac{1}{2}U_1 .
\end{gather}
For a typical energy excitation of $\Delta E=0.5$~meV between the two lowest orbital states in a GaAs quantum dot ($m^*=0.067m_e$), we obtain $R\simeq 34$~nm.
We also use $L\simeq 200$~nm, as consistent with Ref.~\cite{Medford2012}.
We can then estimate the ratios $U_0:U_1:U_2$, which are given by $1:0.1:0.05$.}

\red{
The main effect of including the terms corresponding to $U_1$ and $U_2$ is to suppress the filling of the doubly-occupied states $\ket{4}$ and $\ket{5}$, relative to the doubly-occupied states $\ket{3}$ and $\ket{6}$.
This has a relatively small effect on our numerical results, which we confirm by elimating states $\ket{4}$ and $\ket{5}$ and repeating the analysis.
On the other hand, the model of Eq.~(\ref{eq:HU}) includes fewer parameters, making it more intuitively practical.
The main conclusions of our calculation remain unchanged.}

\section{Exchange Interactions and the Sweet Spot}
In this section, we estimate the effective exchange interactions $J_{ij}$ that generate rotations.
We can use the results to provide initial estimates for the evolution periods for gate operations; we use these estimates to optimize the gates, as discussed in the main text.

We now reduce the full $7\times 7$ Hamiltonian, $H = H_t + H_U + H_\varepsilon$, to an effective $2\times 2$ Hamiltonian for the logical qubit states.
We consider the ideal case with no nuclear fields, so $\Delta B_j=0$ and there is no coupling between the qubit states $\{\ket{0},\ket{1}\}$ and the leakage state $\ket{2}$.
A Schrieffer-Wolff transformation~\cite{SchriefferWolff} to order $t^2$ in the small parameter $t/U$ yields the well-known Heisenberg Hamiltonian
\begin{equation}
H_\text{eff}= J_{12}\,{\bf s}_1\cdot {\bf s}_2+ J_{23}\,{\bf s}_2\cdot {\bf s}_3 , 
\end{equation}
for the $3\times 3$ subspace of $(1,1,1)$ charge states.
Here, $J_{12},J_{23}\sim {\cal O}[t^2]$.
In the absence of any coupling to the leakage state, we can immediately project $H_\text{eff}$ onto the $2\times 2$ logical qubit subspace, yielding
\begin{equation}
H_\text{eff}= \text{(const)} + \frac{\sqrt{3}}{4}(J_{12}-J_{23})\sigma_x-\frac{1}{4}(J_{12}+J_{23})\sigma_z, 
\label{eq:Heff0}
\end{equation}
where $\sigma_x$ and $\sigma_z$ are Pauli matrices.
Here, we may drop the constant term, giving Eq.~(2) in the main text.

We can also obtain $H_\text{eff}$ by directly performing a Schrieffer-Wolff transformation of $H$ onto the $2\times 2$ subspace, yielding
\begin{equation}
H_\text{eff}\simeq -\frac{2\sqrt{3}t^2U\varepsilon_M\varepsilon}{D}\sigma_x
-\frac{2t^2U(U^2-\varepsilon_M^2-\varepsilon^2/4)}{D}\sigma_z , \label{eq:Heff}
\end{equation}
where the denominator is given by
\begin{equation}
D=U^4-2(\varepsilon_M^2+\varepsilon^2/4)U^2+(\varepsilon_M^2-\varepsilon^2/4)^2 .
\end{equation}
By comparing Eqs.~(\ref{eq:Heff0}) and (\ref{eq:Heff}), we can identify the individual exchange interactions:
\begin{gather}
J_{12}=\frac{4t^2U(U^2-(\varepsilon_M+\varepsilon/2)^2)}{D} , \\
J_{23}=\frac{4t^2U(U^2-(\varepsilon_M-\varepsilon/2)^2)}{D} .
\end{gather}

Diagonalizing Eq.~(\ref{eq:Heff}), we obtain the energy splitting
\begin{equation}
E_{01}=\frac{4t^2U\sqrt{(U^2-\varepsilon_M^2-\varepsilon^2/4)^2+3\, \varepsilon_M^2\varepsilon^2}}{D} .
\end{equation}
Since $E_{01}$ is an even function in the variables $\varepsilon$ and $\varepsilon_M$, we immediately find that
\begin{equation}
\frac{\partial E_{01}}{\partial \epsilon} = \frac{\partial E_{01}}{\partial \epsilon_M} = 0 
\end{equation}
when $\epsilon=\epsilon_M=0$, establishing this setting as a detuning sweet spot.

At the sweet spot, we find that
\begin{equation}
H_\text{eff}= -\frac{2t^2}{U} \sigma_z ,
\end{equation}
corresponding to a $Z$-rotation.
Indeed, we see that $Z$-rotations are achieved when either $\varepsilon=0$ or $\varepsilon_M=0$.
From Eq.~(\ref{eq:Heff}), we see that rotations around the axis $-(\hat{\bf x}+\hat{\bf z})/\sqrt{2}$, used in the three-step $X(\pi)$ protocol described in the main text, are defined by the line
\begin{equation}
\varepsilon_M^2+\varepsilon^2/4+\sqrt{3}\, \varepsilon_M\varepsilon = U^2 ,
\end{equation}
which correctly predicts the line of highest fidelities in Fig.~2(b) of the main text.

\section{Quantum Process Tomography}
Quantum process tomography (QPT) provides a means of characterizing quantum gates by comparing the ideal outcomes of gate operations with their actual outcomes.
Here, we follow the QPT recipe given in~\cite{Nielsen&Chuang}.
We solve the master equation, Eq.~(3) in the main text, for a specified pulse sequence for a given gate operation.
For each simulation, the detuning parameters and the local magnetic fields are held constant.
Using the simulation results, we calculate the final fidelity, as outlined below.
In the following section, we describe our method for performing statistical averages of those fidelities, taking into account the fluctuations of the detuning parameters and the random magnetic fields.
We now summarize the QPT method.

We consider a gate operation ${\cal E}(\rho)$ acting on an initial state described by the density matrix $\rho$.
Here, ${\cal E}(\rho)$ represents the final density matrix, and has no relation to the detuning parameter.
The $\cal E$ operation can be expressed in terms of operation elements $E_i$, such that
\begin{equation}
{\cal E} (\rho) = \sum_i E_i\rho E_i^\dagger.
\end{equation}
The operation elements can be decomposed with respect to an orthogonal basis set of operators $\tilde{E}_m$ for the state space, such that
\begin{equation}
E_i=\sum_m e_{im}\tilde{E}_m,
\end{equation}
where $e_{im}$ are complex numbers.
If $\rho$ describes a single qubit, then each $\tilde{E}_m$ is a $2\times 2$ matrix.
A convenient choice is the basis set 
\begin{gather}
\tilde{E}_0 = I ,\\
\tilde{E}_1 = \sigma_x ,\\
\tilde{E}_2 = -i\sigma_y ,\\
\tilde{E}_3 = \sigma_z ,
\end{gather}
where $\sigma_\alpha$ are Pauli matrices.
We then have 
\begin{equation}
{\cal E}(\rho) = \sum_{mn}\tilde{E}_m\rho \tilde{E}_n^\dagger\chi_{mn},
\end{equation}
where the process matrix $\chi$ is defined as
\begin{equation}
\chi_{mn} = \sum_i e_{im}e_{in}^*.
\end{equation}
The process matrix can be fully characterized by initializing the system into linearly independent basis elements for the density matrix.
A convenient choice of initial states is $\ket{0}$, $\ket{1}$, $\ket{+}=(\ket{0}+\ket{1})/\sqrt{2}$, and $\ket{-}=(\ket{0}+i\ket{1})/\sqrt{2}$.
We then perform appropriate linear combinations of gate operations on the initial states, as described in~\cite{Nielsen&Chuang}.
Once the process matrix has been reconstructed, the process fidelity for a single-qubit rotation is given by~\cite{Nielsen2002}
\begin{equation}
\bar{F} = \frac{1}{3}\left(2\text{Tr}[\chi\chi_\text{ideal}]+1\right)  ,
\end{equation}
where $\chi_\text{ideal}$ represents the ideal process matrix.

\section{Averaging Procedure for Overhauser Field and Detuning Fluctuations}
In the previous section, we described the calculation of QPT fidelities for individual simulations.
Each simulation is performed for a constant value of the detuning parameters and the local nuclear fields.
However, these parameters are all quasistatic, and we should perform an average over these quantities, as described in the main text, to describe the inhomogeneous broadening.

There are two different fluctuation axes for the detuning parameters ($\varepsilon$ and $\varepsilon_M$) and two different axes for the nuclear fields ($\Delta B_l$ and $\Delta B_r$).
While it is not computationally feasible to perform accurate, simultaneous averages over four different fluctuation axes, it is possible to perform simultaneous averages over two axes at a time.
We choose to perform simultaneous averages over the detuning fluctuations and the random Overhauser fields separately, to distinguish the effects of charge and nuclear noise.
These calculations are computationally intensive.

We first consider the quasistatic random Overhauser fields.
When $B\gg \Delta B$, we only need to consider the longitudinal components of $\Delta B_{l,r}$~\cite{Taylor2007}.
We model the probability distributions of these random fields as
\begin{equation}
P(\Delta B_l, \Delta B_r) = \frac{1}{2\pi\sigma_B^2}e^{-(\Delta B_l^2 + \Delta B_r^2)/(2\sigma_B^2)},
\end{equation}
where $\sigma_B$ is the standard deviation of the random fields.
The master equation is solved over a grid $(\Delta B_l,\Delta B_r)$ of size $N_g\times N_g$, with $N_g=25$, while keeping $\varepsilon$ and $\varepsilon_M$ fixed.
The noise-averaged fidelity is then given by
\begin{widetext}
\begin{align}
F &= \int\frac{d\Delta B_ld\Delta B_r}{2\pi\sigma_B^2}\bar{F}(\Delta B_l, \Delta B_r,\varepsilon,\varepsilon_M)
e^{-(\Delta B_l^2 + \Delta B_r^2)/(2\sigma_B^2)}\\
  &= \frac{(\Delta B_\text{max} - \Delta B_\text{min})^2}{2\pi\sigma_B^2N_g^2}
  \sum_{\langle \Delta B_l,\Delta B_r \rangle}
  \bar{F}(\Delta B_l, \Delta B_r,\varepsilon,\varepsilon_M)
  e^{-(\Delta B_l^2 + \Delta B_r^2)/(2\sigma_B^2)} .
\end{align}
\end{widetext}
In our simulations, we choose $\Delta B_l$ and $\Delta B_r$ in the range (-12~mT,+12~mT), and $\sigma_B$=4~mT, as consistent with \cite{Assali2011}.

Similarly, we consider fluctuations of the detuning parameters keeping the local magnetic fields fixed.
We model the fluctuation probability distribution as
\begin{equation}
P(\Delta \varepsilon, \Delta \varepsilon_M) = \frac{1}{2\pi\sigma_\varepsilon^2}e^{-(\Delta \varepsilon^2 + \Delta \varepsilon_M^2)/(2\sigma_\varepsilon^2)}
\end{equation}
over a grid $(\Delta \varepsilon,\Delta \varepsilon_M)$ of $N_g\times N_g$ points, with $N_g=31$.
The noise-averaged fidelity is then given by
\begin{widetext}
\begin{align}
F &= \int\frac{d\Delta \varepsilon d\Delta \varepsilon_M}{2\pi\sigma_\varepsilon^2}
\bar{F}(\varepsilon+\Delta \varepsilon, \varepsilon_M+\Delta \varepsilon_M,\Delta B_l,\Delta B_r)
e^{-(\Delta \varepsilon^2 + \Delta \varepsilon_M^2)/(2\sigma_\varepsilon^2)}\\
  &= \frac{(\Delta \varepsilon_\text{max} - \Delta \varepsilon_\text{min})^2}{2\pi\sigma_\varepsilon^2N_g^2}
  \sum_{\langle \Delta \varepsilon,\Delta \varepsilon_M \rangle} 
  \bar{F}(\varepsilon+\Delta \varepsilon, \varepsilon_M+\Delta \varepsilon_M,\Delta B_l,\Delta B_r)
  e^{-(\Delta \varepsilon^2 + \Delta \varepsilon_M^2)/(2\sigma_\varepsilon^2)} .
\end{align}
\end{widetext}
In our simulations, we choose $\Delta \varepsilon$ and $\Delta \varepsilon_M$ in the range (-15~$\mu$eV,+15~$\mu$eV), and $\sigma_\varepsilon$=5~$\mu$eV, as consistent with~\cite{Petersson2010,Shi2012}.

\section{Nuclear Noise Averages}
Figures 2(a) and (c) of the main text show comparisons of nuclear and charge noise averaging results, while the insets show comparisons of charge noise averaging for three different values of $U$.

In this section, we extend these results by plotting nuclear noise-averaged results for three different values of $U$, as shown in Fig.~S1.
As before, we find that $X(\pi)$ rotations have fidelities that are approximately 20 times worse than $Z(\pi)$ rotations, with optimal values that improve when $U$ is larger.

\section{$^{28}$Si}
In previous sections, particularly in Fig. 2 of the main text and Fig.~S1 of the Supplemental Materials, we compared the effects of random nuclear fields and detuning fluctuations. 
When we simulated detuning fluctuations, we adopted a fixed, characteristic magnetic field difference between the two dots.
To complete this story, we perform the same simulation here, setting the static Overhauser fields to zero.
This can be viewed as the ideal case for perfect, isotopically purified $^{28}$Si devices, whereas the previous simulations corresponded to GaAs.

The results of our $^{28}$Si simulation are shown in Fig.~S2, assuming only detuning fluctuations.
We also show the equivalent GaAs simulation for comparison, with the same detuning fluctuations but setting $\Delta B_2^z-\Delta B_1^z=\Delta B_3^z-\Delta B_2^z =3$~mT, as in Fig.~2 of the main text.
We see that using $^{28}$Si is highly beneficial in two ways.  
First, it yields an improvement in the maximum fidelity.
Second, it lowers the optimal tunnel coupling, and therefore the gate speed, to a range that may be more convenient from a technological perspective.
At even lower gate speeds, fast charge noise eventually degrades the gate fidelity.
At higher tunnel couplings, where the effects of quasistatic charge noise and leakage to doubly-occupied charge states dominate the fidelity, the presence of a magnetic field difference is irrelevant.
Once again, we find the optimal fidelity for $Z$-rotations is approximately one order of magnitude better than for $X$-rotations, due to the presence of the sweet spot.

\end{appendix}

\end{document}